# Airborne Radar STAP using Sparse Recovery of Clutter Spectrum

Ke Sun, Hao Zhang, Gang Li, Huadong Meng, Xiqin Wang

(Department of Electronic Engineering, Tsinghua University, Beijing 100084, China)

**Abstract**: Space-time adaptive processing (STAP) is an effective tool for detecting a moving target in spaceborne or airborne radar systems. Statistical-based STAP methods generally need sufficient statistically independent and identically distributed (IID) training data to estimate the clutter characteristics. However, most actual clutter scenarios appear only locally stationary and lack sufficient IID training data. In this paper, by exploiting the intrinsic sparsity of the clutter distribution in the angle-Doppler domain, a new STAP algorithm called SR-STAP is proposed, which uses the technique of sparse recovery to estimate the clutter space-time spectrum. Joint sparse recovery with several training samples is also used to improve the estimation performance. Finally, an effective clutter covariance matrix (CCM) estimate and the corresponding STAP filter are designed based on the estimated clutter spectrum. Both the Mountaintop data and simulated experiments have illustrated the fast convergence rate of this approach. Moreover, SR-STAP is less dependent on prior knowledge, so it is more robust to the mismatch in the prior knowledge than knowledge-based STAP methods. Due to these advantages, SR-STAP has great potential for application in actual clutter scenarios.

## 1. Introduction

An airborne/spaceborne (A/S) space time adaptive processor (STAP) attempts to detect a moving target in the presence of a Doppler/angle spread clutter environment [1-2]. Due to the motion of the radar platform, one dimensional processing in neither angle nor Doppler domain can effectively distinguish the moving target from the surrounding clutter environment. Therefore, it is necessary to carry out joint angle-Doppler processing. The fundamental component of STAP is the effective estimation of the clutter covariance matrix (CCM), which is used to construct the optimal linear weighting of the adaptive matched filter such that the output signal-to-clutter ratio is maximized [1]. The adaptive processor requires a certain quantity of independent identically distributed (IID) training samples to effectively estimate CCM due to lack of knowledge about the external clutter environment [3-4]. The number of IID samples required to produce an output signal-to-clutter power ratio (SCR) performance that is close to the optimum (nominally, within 3 dB) is called the convergence rate of the processor. Reducing the convergence rate is quite meaningful because the clutter is locally stationary in the actual scenario and the CCM estimation with slow convergence rate will cause a significant SCR loss between the adaptive implementation and the optimum design [5].

The foundational work by Reed, called sample matrix inversion (SMI) [3], provides a way to estimate the CCM. This method uses data samples directly and the convergence rate is twice the problem dimensionality. However, in actual clutter scenarios, this quantity of IID samples is often unavailable due to the nonstationarity of the reflectivity properties, strong discrete scatters, and so forth [5-6]. Thus, faster-converging methods, such as loaded sample matrix inversion (LSMI) and principal component (PC), are proposed in [6-7]. These methods avoid the use of eigenvectors associated with small eigenvalues that are difficult to estimate. In this way, the convergence rate is reduced to twice the number of significant eigenvalues (referred to as the 'effective clutter rank') of the CCM. However, the convergence rate can hardly be improved beyond the clutter rank because these methods do not assume any prior knowledge in the CCM estimation.

Recently, a class of algorithms called KB-STAP (knowledge-based STAP) [8-12] was proposed to solve the problem of lacking IID training samples. These methods use prior knowledge of the spectral characteristics in the underlying clutter scenario to pre-whiten the CCM and enhance the convergence rate. This prior knowledge includes the radar parameters and anticipated structure of the clutter return, which could be obtained from a digital terrain and elevation data map, synthetic aperture radar (SAR) imagery, and some other sensors [8]. Two important KB-STAP algorithms are colored loading (CL) [9] and fast maximum likelihood with assumed clutter covariance (FMLACC) [10]. It has been proved that both approaches are equivalent to a kind of pre-whitening process and the convergence rate is twice the rank of the pre-whitened CCM. However, it should also be noted that accurate prior knowledge is hard to acquire due to the limited computational resources, insufficient resolution, inadequate array calibration and so forth [9-10]. Therefore, the mismatch between the assumed and the actual CCM will impact the performance of the KB-STAP methods.

Focusing on the problem of slow convergence rate in the conventional STAP and sensitivity to the prior knowledge in the KB-STAP methods, a new STAP algorithm is proposed in this paper. Similar to the idea in KB-STAP, this method also utilizes prior knowledge to obtain the clutter spectral characteristics. However, this spectrum estimation is obtained via sparse recovery, which has evolved very rapidly in the last decade [13-18]. In its most basic form, the problem of sparse recovery attempts to find the sparsest signal $\boldsymbol{\alpha}$ to satisfy $\mathbf{x} = \boldsymbol{\Psi}\boldsymbol{\alpha}$, where $\boldsymbol{\Psi} \in C^{m \times n}$ is an overcomplete basis, i.e., m ≤ n. Without prior knowledge that $\boldsymbol{\alpha}$ is sparse, the equation $\mathbf{x} = \boldsymbol{\Psi}\boldsymbol{\alpha}$ is ill-posed and has many solutions. Additional information that $\boldsymbol{\alpha}$ should be sufficiently sparse allows one to eliminate this ill-posedness [13-14]. Solving the ill-posed problem involving sparsity typically requires combinatorial optimization, which is intractable even for modest data size. A number of practical algorithms such as convex optimization (including $L_1$ norm minimization) [15-16] and greedy algorithms [17-18] have been proposed to approximate the solution to this problem. Convex optimization provides a guarantee of uniformity and is also very stable. However, this method is based on linear programming and has a quite high computational complexity, $O(n^3)$. On the other hand, greedy methods make a sequence of locally optimal choices in an effort to obtain a globally optimal solution. The computational load is quite small, but it lacks the strong guarantee of convergence that convex optimization provides. Prior research has proved sparse recovery as a valuable tool for spectrum estimation, but its application has been mainly focused in the field of source localization, where the sparsity condition is obviously satisfied [19-21].

In this paper, we exploit the prior knowledge of the sparsity of the clutter distribution in the angle-Doppler domain and obtain the space-time spectrum via sparse recovery. This method can obtain the clutter spectral characteristics with many fewer IID samples, such that the convergence rate of the CCM is significantly enhanced. The remainder of this paper is organized as follows. Section 2 describes the basic model of the STAP received data. In section 3, the SR-STAP (STAP via sparse recovery) approach is proposed to estimate the clutter space-time spectrum from both single and multiple snapshots. Then, the CCM estimation based on this spectrum is given to effectively suppress the clutter. Section 4 analyzes the convergence rate and robustness for simulated data. The Mountaintop data [22] provided by Lincoln Laboratories is also used to verify the effectiveness of SR-STAP in an actual clutter scenario. Section 5 presents concluding remarks about the proposed algorithm and points out directions for future work.

## 2. STAP MODEL

When a radar platform is moving, stationary clutter has an angle-Doppler dependence given by

$$f_d = \frac{2v}{\lambda}\sin\theta_s, \qquad (1)$$

where $\theta_s$ and $f_d$ stand for the spatial angle and Doppler frequency of the clutter scatter, respectively; $v$ is the moving velocity of the radar platform; $\lambda$ denotes the radar wavelength. Suppose that $N$ is the number of array channels and $M$ is the number of pulses in a coherent process interval (CPI). The target-free snapshot $x \in \mathrm{C}^{NM}$ for a given range cell can be expressed as [1]

$$\mathbf{x} = \sum_{i=1}^{N_c} \gamma_i \cdot \boldsymbol{\varphi}(\theta_{s,i}, f_{d,i}) + \mathbf{n}, \qquad (2)$$

where $\boldsymbol{\varphi}(\theta_{s,i}, f_{d,i})$ is the $NM \times 1$ space-time steering vector of the $ith$ clutter scatter, given by

$$\begin{aligned}\boldsymbol{\varphi}(\theta_{s,i}, f_{d,i}) = &\left[1,\ \exp\left(j2\pi\frac{f_{d,i}}{\mathrm{PRF}}\right),...,\exp\left(j2\pi(M-1)\frac{f_{d,i}}{\mathrm{PRF}}\right)\right]^T \\ &\otimes \left[1,\ \exp\left(j2\pi\frac{d}{\lambda}\sin\theta_{s,i}\right),...,\exp\left(j2\pi(N-1)\frac{d}{\lambda}\sin\theta_{s,i}\right)\right]^T,\end{aligned} \qquad (3)$$

$\theta_{s,i}, f_{d,i}$ denote the space angle and Doppler frequency of the $ith$ clutter scatter, $\otimes$ denotes the Kronecker product, PRF is the radar pulse repetition frequency, $d$ is the inter-sensor spacing of the uniform linear array. $N_c$ is the total number of the uncorrelated clutter scatters, $\mathbf{n} \sim CN(\mathbf{0}, \delta^2\mathbf{I})$ is the complex gauss noise ($\mathbf{I}$ is the $NM \times NM$ identity matrix), and $\gamma_i$ is the complex amplitude of the $ith$ clutter scatter satisfying

$$E\{\gamma_i\} = 0,\ \forall 1 \le i \le N_c, \qquad (4)$$

The optimal STAP is to design the $NM \times 1$ space-time filter to maximize the output SCR and the corresponding optimal space-time filter is given as [1-2]

$$\mathbf{w}_{opt} = \mu \mathbf{R}^{-1} \mathbf{s}, \tag{5}$$

where $\mu$ is a non-zero constant, matrix $\mathbf{R} = E\{\mathbf{xx}^H\}$ represents the actual CCM, and $\mathbf{s}$ is the space-time steering vector of the moving target. In practice, the CCM is unknown and should be estimated using IID snapshots. In the SMI approach [3], the CCM estimation is given via the direct data approach:

$$\hat{\mathbf{R}}_{SMI} = \frac{1}{L} \sum_{i=1}^{L} \mathbf{x}_i \mathbf{x}_i^H, \tag{6}$$

where $\mathbf{x}_i, 1 \leq i \leq L$ denote the IID snapshots. This approach can achieve nearly optimal performance when the number of IID snapshots satisfies $L \geq 2MN$. However, this quantity of stationary samples is often unavailable in an actual clutter scenario [5]. One great improvement to this method is called LSMI (Loaded SMI), which adds a small diagonal loading factor into the SMI estimation:

$$\hat{\mathbf{R}}_{LSMI} = \hat{\mathbf{R}}_{SMI} + \beta_L \mathbf{I} \tag{7}$$

where $\beta_L$ is a small loading factor. It has been proved that this approach can achieve nearly optimal performance using twice as many IID samples as the effective clutter rank [6]. However, the convergence rate can hardly be improved beyond the effective clutter rank because it does not use any prior knowledge in the CCM estimation. At the same time, substituting (2) into the CCM definition, the CCM can also be expressed as [1]

$$\mathbf{R} = E[\mathbf{xx}^H] = E\left\{\left(\sum_{i=1}^{N_c} \gamma_i \boldsymbol{\varphi}(\theta_{s,i}, f_{d,i}) + \mathbf{n}\right)\left(\sum_{j=1}^{N_c} \gamma_j \boldsymbol{\varphi}(\theta_{s,j}, f_{d,j}) + \mathbf{n}\right)^H\right\}. \tag{8}$$

The clutter scatters are uncorrelated with each other, that is

$$E\{\gamma_i \gamma_j^*\} = E\{\gamma_i\} E\{\gamma_j^*\} = 0, \forall i, j : i \neq j. \tag{9}$$

Moreover, the noise is uncorrelated with these clutter scatters. In this case, the CCM could be further simplified as

$$\mathbf{R} = \sum_{i=1}^{N_c} E\{|\gamma_i|^2\} \boldsymbol{\varphi}(\theta_{s,i}, f_{d,i}) \boldsymbol{\varphi}(\theta_{s,i}, f_{d,i})^H + \delta^2 \mathbf{I}, \tag{10}$$

where $E\{|\gamma_i|^2\}, 1 \leq i \leq N_c$ denotes the power distribution of the clutter scatters in the angle-Doppler domain, that is, the space-time spectrum. In this way, the relationship between the CCM and the clutter spectral characteristics is built up. The assumed CCM can be used as an approximation for the actual one, effectively accelerating the convergence rate, as long as the prior knowledge of the clutter distribution is obtained accurately. The recently-developed knowledge-based STAP methods are based on this idea [8-12]. By incorporating the prior knowledge into the estimation process, these methods show a strong capability to reduce the number of required data samples. One important knowledge-based STAP algorithm is called CL [9], which provides the following estimate for the CCM:

$$\hat{\mathbf{R}}_{CL} = \hat{\mathbf{R}}_{SMI} + \beta_d \mathbf{R}_c + \beta_L \mathbf{I} \qquad (11)$$

where $\hat{\mathbf{R}}_{SMI}$ is the SMI part standing for the contribution of the data samples, $\mathbf{R}_c$ is the assumed CCM using the prior knowledge in (10), $\beta_d$ and $\beta_L$ denote the colored loading and diagonal loading factors respectively. The CL method utilizes the prior knowledge of the clutter distribution to force the nulls in the angle-Doppler domain and effectively suppress the corresponding clutter. In this way, the convergence rate improves to twice the effective rank of the pre-whitened clutter residual, which is not contained in the prior knowledge. However, accurate prior knowledge is hard to acquire due to various real-world effects, such that there will be some performance degradation in KB-STAP due to the mismatch between the prior knowledge and the real-world scenario [9-10].

## 3. Spectrum Estimation and Clutter Suppression

As stated above, the CCM estimation could be obtained with high performance as long as the clutter spectral characteristics are accurately acquired. Based on this idea, a new STAP algorithm is developed via sparse recovery to obtain the clutter spectrum and estimate the CCM with much less IID snapshots [23]. The space angle and Doppler frequency axes are discretized into $N_s = \rho_s N, N_d = \rho_d M$ grids in the angle-Doppler domain to obtain the clutter spectrum with high resolution. The parameters $\rho_s, \rho_d$ are the resolution scales along the angle and Doppler axes, respectively. In this way, the received data in (2) can be written in matrix form as

$$\mathbf{x} = \sum_{i=1}^{N_s N_d} \alpha_i \cdot \mathbf{\Psi}_i + \mathbf{n} = \mathbf{\Psi}\boldsymbol{\alpha} + \mathbf{n}, \qquad (12)$$

where $NM \times N_s N_d$ matrix $\mathbf{\Psi}$ is the basis made up with the space-time steering vectors and can be expressed as

$$\mathbf{\Psi} = \left[ \boldsymbol{\varphi}(\theta_{s,1}, f_{d,1}), \cdots, \boldsymbol{\varphi}(\theta_{s,N_s}, f_{d,1}), \cdots, \boldsymbol{\varphi}(\theta_{s,1}, f_{d,N_d}), \cdots, \boldsymbol{\varphi}(\theta_{s,N_s}, f_{d,N_d}) \right]. \quad (13)$$

The vector $\boldsymbol{\alpha}$ stands for the clutter distribution in the basis $\mathbf{\Psi}$, that is, the clutter space-time spectrum. Equation (12) is the fundamental equation in this paper and has two characteristics that we should pay attention to. First, estimating the space-time spectrum $\boldsymbol{\alpha}$ is equivalent to solving the linear equation (12) with the data $\mathbf{x}$. Second, the basis matrix $\mathbf{\Psi}$ is overcomplete and the problem is ill-posed because the resolution scales $\rho_s, \rho_d$ are greater than the one used to obtain the high-resolution spectrum. However, according to the theory of sparse recovery [13-16], when the actual clutter spectral distribution $\boldsymbol{\alpha}_0$ is sparse, this ill-posed problem can be solved via sparse recovery. Next, some discussion is given to verify the sparse characteristic of the space-time clutter in the discretized angle-Doppler plane.

### 3.1 The sparsity of the space-time clutter

As stated above, the angle-Doppler domain is discretized into $N_s, N_d$ cells along the angle and Doppler axes, respectively. Each cell in this discretized plane corresponds to a certain space-time steering vector and all these vectors make up the overcomplete basis $\mathbf{\Psi}$. Because the noise does not correspond to a certain space-time steering vector, its distribution is reflected as a small noise floor in the angle-Doppler plane. As shown in Fig. 1, the angle-Doppler dependence in (1) focuses the clutter distribution only along the clutter ridge, whose slope is determined by the radar parameters. The STAP clutter scenario usually has a high CNR [1-2], therefore the distribution of the training data in the angle-Doppler plane is mainly determined by the clutter component. Moreover, the clutter scatters from the sidelobe are much smaller (more than 10dB lower) than that from the mainlobe due to the effect of antenna azimuth weighting. In this way, the significant clutter scatters only exist along the clutter ridge within the mainlobe $[\theta_{\min}, \theta_{\max}]$. Conventionally, the significant elements of the actual solution are used to express the degree of sparsity in the sparse recovery [13-14]. In our problem of spectrum estimation, the sparsity of the space-time clutter is equal to the number of cells occupied by the clutter ridge in the discretized angle-Doppler plane, which is marked by the slash cells in Fig. 1.

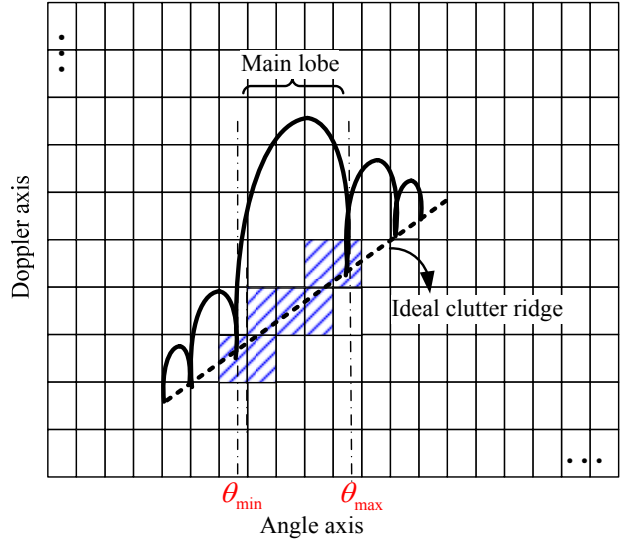

Fig. 1 Clutter distribution in the discretized angle-Doppler plane

In an actual clutter scenario, the sparsity of the actual clutter distribution is unknown and need to be estimated. Similar to the knowledge-based methods, the assumed sparsity could be similarly given by the number of cells occupied by the assumed clutter ridge based on the prior knowledge. Generally, the sparsity is related to both the actual clutter distribution and the number of the discretized cells in the angle-Doppler plane. Due to the effect of the discretization, the sparsity estimation in the angle-Doppler plane may differ from that in the continuous case. An effective estimation method is given to obtain the sparsity as follows.

1. Determine the azimuth angle cells occupied by the clutter distribution as

$$\Delta N = \left\lceil \frac{(\theta_{\min} - \theta_{\max})}{180} \cdot \alpha_s N \right\rceil, \tag{14}$$

   where $[\theta_{\min}, \theta_{\max}]$ stands for the mainlobe angle extent and $\lceil \cdot \rceil$ is the ceiling operation (rounding up).

2. Calculate the corresponding Doppler cells due to the clutter spreading as

$$\Delta M = \left\lceil \frac{2v(\sin\theta_{\max} - \sin\theta_{\min})}{\lambda \cdot \text{PRF}} \cdot \alpha_d M \right\rceil. \tag{15}$$

3. Theoretically, sparsity estimation has a range value $\max[\Delta M, \Delta N] \leq \hat{s} \leq \Delta M + \Delta N$. However, when the resolution scales $\rho_s, \rho_d$ are much greater than the one, we can calculate it directly as

$$\hat{s} = \left\lceil \sqrt{\Delta M^2 + \Delta N^2} \right\rceil. \tag{16}$$

This estimation clearly illustrates the relationship between the radar parameters and the sparsity. Although it is simple and may be not very accurate, the simulation in the next section will show that SR-STAP is robust to this sparsity estimation. Some other practical factors, such as crab angle, also affect the assumed sparsity and will be discussed in detail later.

### 3.2 Space-time spectrum estimation via sparse recovery

**A. Single snapshot case**

Once the actual clutter spectrum $\boldsymbol{\alpha}_0$ is sparse, the ill-posed problem in (12) can be solved using the $L_1$ norm minimization, constructed as

$$\hat{\boldsymbol{\alpha}} = \arg\min \|\boldsymbol{\alpha}\|_1 \ \text{subject to} \ \|\mathbf{x} - \boldsymbol{\Psi}\boldsymbol{\alpha}\|_2 \leq \varepsilon, \tag{17}$$

where $\|\cdot\|_p$ stands for the $L_p$ norm and $\varepsilon$ is the error allowance in sparse recovery. In this way, the $L_2$ norm constraint by $\varepsilon$ guarantees the residual $\|\mathbf{x} - \boldsymbol{\Psi}\boldsymbol{\alpha}\|_2$ to be small, whereas the $L_1$ norm enforces the sparsity of the estimated spectrum. The sparse recovery via the $L_1$ norm minimization can be efficiently carried out via convex optimization. Great achievements in sparse recovery [13] have also proved that even in the scenario with small noise, the $L_1$ norm recovery can approximate the actual solution as $\|\hat{\boldsymbol{\alpha}} - \boldsymbol{\alpha}_0\|_2 \leq \Lambda \cdot \varepsilon$, where $\Lambda$ is the stability coefficient and related to the maximal mutual coherence in the matrix $\boldsymbol{\Psi}$.

Next, the Mountaintop data [22] is used to verify the effectiveness of the spectrum estimation in SR-STAP. The Mountaintop program consists of 14 array sensors and 16 pulses in each CPI. As for the clutter scenario, the dominant clutter is located at about azimuth -15 degrees and normalized Doppler frequency 0.25. The clutter in other directions is small due to shadowing by nearer-range mountains [24]. The resolution scales are set as $\rho_s = 4, \rho_d = 4$ to obtain the high-resolution spectrum. The Capon estimator [25] is introduced here as a reference to verify the performance of the SR-STAP spectrum. The number of snapshots is set to 40 (range cells 80-120). In SR-STAP, only one snapshot (range cell 120) is used to estimate the clutter spectrum. Here, cvx, the most popular convex optimization package, is employed as the $L_1$ norm optimization tool

[26]. The spectra estimated by Capon and SR-STAP are shown in Fig. 2 (a) and (b), respectively.

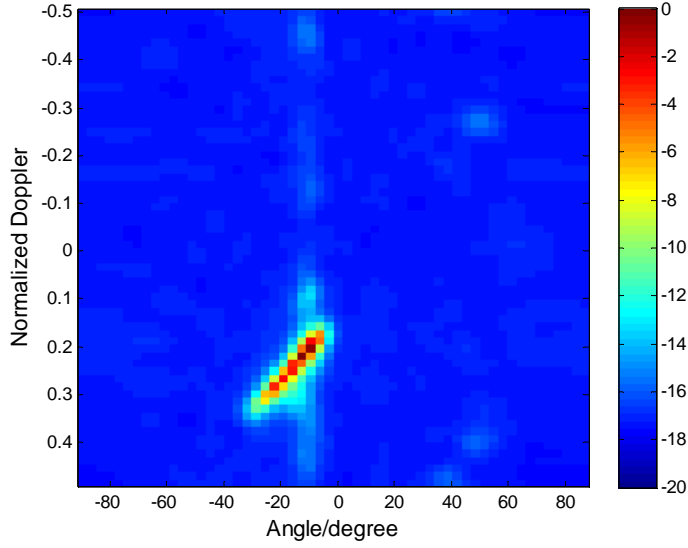

Fig. 2 (a) Capon spectrum (dB) with 40 IID snapshots.

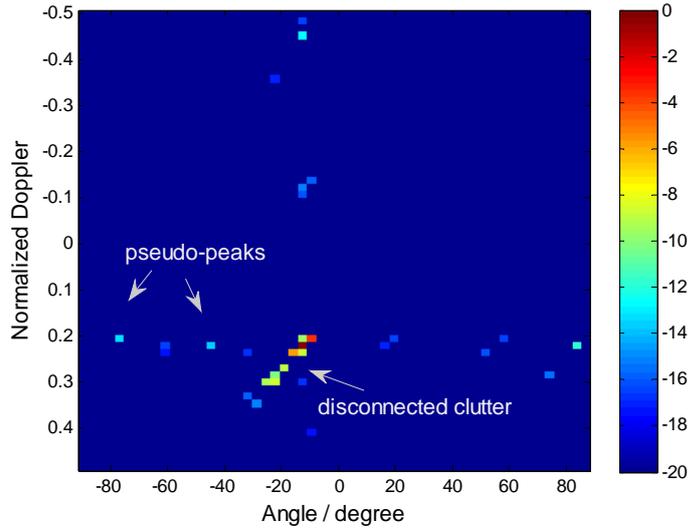

Fig. 2 (b) SR-STAP spectrum (dB) with single snapshot.

It can be seen that most components in the Capon space-time spectrum are very small and only a few dominant scatters are located around the actual clutter area. Meanwhile, SR-STAP approximately estimates the dominant clutter scatters with super-resolution using a single snapshot. However, some estimation discrepancy still exists and needs to be refined. First, the snapshot $\mathbf{x}$ tends to be expressed with fewer space-time steering vectors in the overcomplete basis and may obtain a more sparse solution $\hat{\boldsymbol{\alpha}}$ than the actual scenario due to the characteristics of the $L_1$ norm minimization. In this way, the actual continuous clutter distribution may converge into several disconnected clutter scatters, marked by the ellipse in Fig. 2 (b). This phenomenon is quite common in the field of sparse recovery [19-21]. Second, some estimation error, called 'pseudo-peaks,' exists. These are marked by the arrow in the SR-STAP spectrum. This is caused by the noise and ill-posedness in (12). Normally, these pseudo-peaks appear randomly in the angle-Doppler domain among different snapshots and their amplitudes are much smaller (-10dB) than the actual clutter scatters. The following work will focus on how to eliminate these pseudo-peaks, as well as obtain the connected clutter spectrum with multiple snapshots.

**B. Multiple snapshots case：simple average**

In the actual clutter scenario, although sufficient training samples are hard to guarantee, there are still a few training samples that we can exploit. Therefore, we extend this work to the multiple snapshots case [27-28] to improve the estimation performance as

$$\mathbf{X} = \mathbf{\Psi}\mathbf{S} + \mathbf{N}, \tag{18}$$

where $NM \times L$ matrix $\mathbf{X} = \left[\mathbf{x}^{(1)}, \cdots \mathbf{x}^{(L)}\right]$ denotes the multiple training snapshots, $NM \times L$ matrix $\mathbf{N} = \left[\mathbf{n}^{(1)}, \cdots \mathbf{n}^{(L)}\right]$ is the observation noise, and $N_s N_d \times L$ matrix $\mathbf{S} = \left[\boldsymbol{\alpha}^{(1)}, \cdots \boldsymbol{\alpha}^{(L)}\right]$ is the clutter spectrum of multiple snapshots. A simple method is to separate this joint problem into a series of independent subproblems as

$$\mathbf{x}^{(k)} = \mathbf{\Psi}\boldsymbol{\alpha}^{(k)} + \mathbf{n}^{(k)}, \ 1 \le k \le L. \tag{19}$$

Each subproblem can be solved via the $L_1$ norm minimization in (17) to obtain the sparse spectrum estimation. Then, the average of these estimated spectra $\hat{\boldsymbol{\alpha}}^{(k)}, \ 1 \le k \le L$, can be taken as

$$\hat{\boldsymbol{\alpha}} = \frac{1}{L}\sum_{k=1}^{L}\hat{\boldsymbol{\alpha}}^{(k)}. \tag{20}$$

This method is simple and the computing effort is linearly proportional to the number of snapshots. The estimation result via simple average is shown in Fig. 3 for Mountaintop data range cells 80-120. It is shown that this approach can improve the disconnected spectrum estimation to some extent, compared to the single snapshot method. However, some pseudo-peaks still exist after simple averaging because this method does not make use of the stationary characteristic of the clutter distribution in the IID snapshots. In fact, the position is more important than the amplitude information in the clutter distribution.

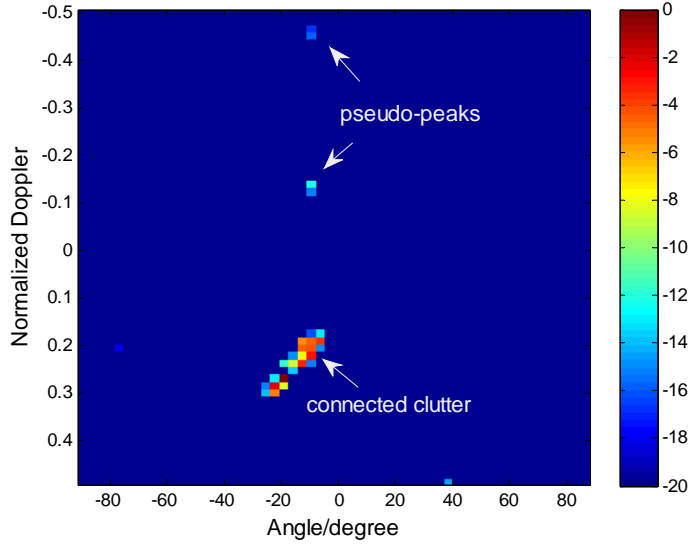

Fig. 3 SR-STAP spectrum (dB) with simple average

## C. Multiple snapshots case: joint sparse recovery

To improve the spectrum performance of the simple average approach, a more reasonable approach called joint sparse recovery [29-30] is proposed to treat these snapshots in conjunction. This assumes that the sparse structure of a stationary process remains the same across different snapshots and the only difference is reflected in the amplitude variation. In this case, the sparsity $s$ in each spectrum $\mathbf{\alpha}^{(k)}$, $1 \leq k \leq L$ is identical, which means that the actual clutter scatters are located at the same cells in the angle-Doppler domain in these spectra $\mathbf{\alpha}^{(k)}$, $1 \leq k \leq L$ as

$$\mathbf{\Gamma}^{(1)} = \cdots \mathbf{\Gamma}^{(k)}, \tag{21}$$

where the position set $\mathbf{\Gamma}^{(k)}$ in the $kth$ snapshot is defined as

$$\mathbf{\Gamma}^{(k)} = \arg \max_s \left( \left| \mathbf{\alpha}^{(k)} \right| \right), \ 1 \leq k \leq L, \tag{22}$$

where operation $\arg \max_s (\cdot)$ denotes obtaining the positions of the maximal $s$ elements in a vector. Prior research on joint sparse recovery achieved great recovery performance with the assumptions of accurate knowledge of the sparsity and small maximal mutual coherence in the overcomplete matrix $\mathbf{\Psi}$. However, these assumptions are not valid in the problem of spectrum estimation in the actual clutter scenario because the sparsity of the actual clutter scenario is unknown and the maximal mutual coherence in the overcomplete matrix $\boldsymbol{\Psi}$ increases with the resolution scales $\rho_s, \rho_d$. For example, when $\rho_s, \rho_d \gg 1$, the maximal mutual coherence may become large, such that the clutter spectrum estimation via the current joint sparse recovery method is not very effective. Therefore, some practical improvements should also be considered to enhance the current methods of joint sparse recovery.

As stated above, the spectrum estimation with the single snapshot is discontinuous along the actual clutter area and may have some pseudo-peaks. Because these snapshots are joint sparse, the combing result with position sets $\hat{\boldsymbol{\Gamma}}^{(k)}, 1 \leq k \leq L$ is more close to the continuous distribution of the actual clutter. Moreover, the pseudo-peaks could be effectively suppressed using this combining operation because they appear randomly in the spectrum estimation of the different snapshots. Thus, the joint problem with multiple snapshots could be effectively solved via the least squares (LS) method as soon as the actual position set is accurately obtained [33]. The procedure for this method is elaborated as follows.

1. Decompose the joint problem into a series of subproblems and solve each subproblem independently as

$$\hat{\boldsymbol{\alpha}}^{(k)} = \arg\min \|\boldsymbol{\alpha}^{(k)}\|_1 \text{ subject to } \|\mathbf{x}^{(k)} - \boldsymbol{\Psi}\boldsymbol{\alpha}^{(k)}\|_2 \leq \varepsilon, \ 1 \leq k \leq L. \tag{23}$$

2. Estimate the position set of the clutter distribution in each space-time spectrum as

$$\hat{\boldsymbol{\Gamma}}^{(k)} = \arg\max_s \left(\left|\hat{\boldsymbol{\alpha}}^{(k)}\right|\right), \ 1 \leq k \leq L. \tag{24}$$

3. Define a series of $N_s N_d \times 1$ vectors $\mathbf{c}^{(1)}, \cdots \mathbf{c}^{(L)}$ as

$$\mathbf{c}^{(k)} = \begin{cases} if \ i \in \hat{\boldsymbol{\Gamma}}^{(k)}, \ c_i^{(k)} = 1 \\ else, \ c_i^{(k)} = 0 \end{cases}, \ 1 \leq i \leq N_s N_d, \ 1 \leq k \leq L. \tag{25}$$

4. Combine these vectors as $\mathbf{P} = \mathbf{c}^{(1)} + \cdots + \mathbf{c}^{(L)}$, with the element $P_i$ standing for the occurrence number on the $ith$ space-time cell after the combining process.

5. Obtain the joint position set with the position sets of multiple snapshots as

$$\hat{\boldsymbol{\Gamma}} = \arg\max_s(\mathbf{P}). \tag{26}$$

6. Once the joint position set $\hat{\boldsymbol{\Gamma}}$ is recovered, the joint problem can be expressed as [24]

$$\min_{\mathbf{S}_{\hat{\Gamma}}} \left\| \mathbf{\Psi}_{\hat{\Gamma}} \mathbf{S}_{\hat{\Gamma}} - \mathbf{X} \right\|_2^2, \tag{27}$$

where $NM \times s$ matrix $\mathbf{\Psi}_{\hat{\Gamma}}$ stands for the corresponding $\hat{\Gamma}$ columns in the overcomplete basis $\mathbf{\Psi}$, then the LS minimization can be effectively implemented to solve the problem as

$$\mathbf{S}_{\hat{\Gamma}} = \left( \mathbf{\Psi}_{\hat{\Gamma}}^H \mathbf{\Psi}_{\hat{\Gamma}} \right)^{-1} \mathbf{\Psi}_{\hat{\Gamma}}^H \mathbf{X}, \tag{28}$$

the $s \times L$ matrix $\mathbf{S}_{\hat{\Gamma}}$ stands for the corresponding $\hat{\Gamma}$ rows in the matrix $\mathbf{S}$ and all other elements of the matrix are set to zero.

7. Calculate the average spectrum of joint sparse recovery as

$$\hat{\boldsymbol{\alpha}} = \frac{1}{L} \sum_{k=1}^{L} \mathbf{S}_k, \ 1 \leq k \leq L. \tag{29}$$

where $\mathbf{S}_k$ stands for the *kth* column of the matrix $\mathbf{S}$ and $\hat{\boldsymbol{\alpha}}$ represents the final spectrum estimation with multiple snapshots.

Until now, we have introduced the SR-STAP spectrum based on joint sparse recovery. The result in Fig. 4 shows that most of the pseudo-peaks are effectively suppressed, which can be explained as follows. There are fewer false positions induced by the pseudo-peaks in the joint position set $\hat{\Gamma}$ because the position information is more reliable than the amplitude information. Moreover, even though some false positions are contained in the final position set $\hat{\Gamma}$, the projection coefficients onto them are quite small because LS has projected most of the measurement $\mathbf{X}$ onto the actual subspace $\mathbf{S}_\Gamma$ to minimize the $L_2$ norm distance. Therefore, the final spectrum estimate is not sensitive to the impact of the pseudo-peaks. It can be concluded that joint sparse recovery could reduce the impact of the pseudo-peaks and obtain a connected spectrum estimate in the actual clutter area.

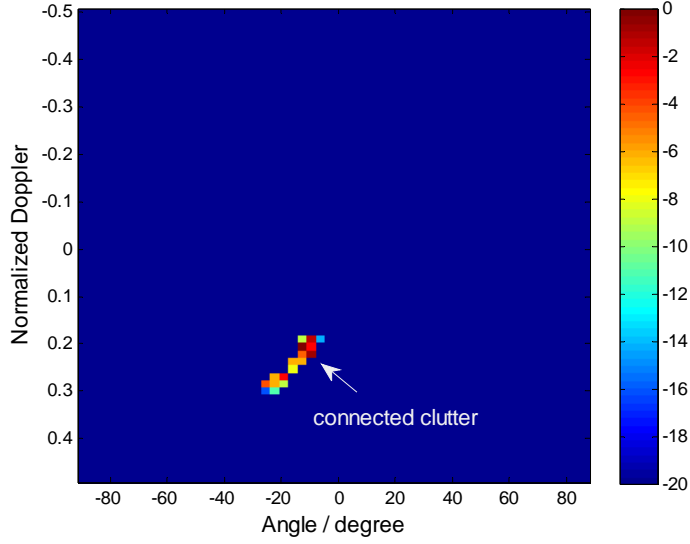

Fig. 4 SR-STAP spectrum (dB) with the joint sparse recovery

**3.2 Covariance matrix estimation**

Once SR-STAP can estimate the clutter distribution in the angle-Doppler domain accurately, the CCM estimation can be given as

$$\hat{\mathbf{R}}_{SR} = \sum_i |\hat{\alpha}_i|^2 \boldsymbol{\varphi}(\theta_{s,i}, f_{d,i}) \boldsymbol{\varphi}(\theta_{s,i}, f_{d,i})^H + \beta_L \mathbf{I}, \quad (30)$$

where $|\hat{\alpha}_i|^2$ is the space-time power spectrum for the $ith$ clutter scatter and $\beta_L$ is a small loading factor. Most elements in $|\hat{\boldsymbol{\alpha}}|^2$ are quite small and there are only few significant elements standing for the influence of the actual clutter scatters $E\{|\gamma_i|^2\}, 1 \leq i \leq N_c$. Finally, the space-time adaptive filter in SR-STAP can be given as

$$\mathbf{w}_{SR} = \mu \hat{\mathbf{R}}_{SR}^{-1} \mathbf{s}. \quad (31)$$

## 4. Experimental Results

In this section, the target detection performance is first shown for Mountaintop data to confirm the effectiveness of SR-STAP. Next, some simulations are presented to test the convergence rate and robustness of different STAP approaches.

**4.1 Target detection performance**

In the Mountaintop program [24], a synthetic target is introduced at the location of range cell 146, azimuth 15 degrees and normalized Doppler 0.25. As shown by the 'Non' line in Fig. 5 (a), the target is completely obscured by the sidelobe leakage from the dominant clutter area without adaptive beamformer processing because the dominant clutter and the target have the same Doppler frequency. Fig. 5 (a) gives the output power along the range cell with 6 training samples for both LSMI and SR-STAP (the nearest 4 cells around the test cell are guard cells and not included in the training samples). It is shown that SR-STAP can effectively suppress the dominant clutter and make the maximum clutter residual 8 dB below the actual target. For comparison, LSMI with the same amount of samples can hardly obtain a sufficient estimate of the actual clutter distribution. Thus, there is still strong residual clutter after the adaptive processing and the target is still submerged in the surrounding clutter environment. Fig. 5 (b) gives the corresponding results with 12 training samples, where the performance of both SR-STAP and LSMI improves to some extent. However, this quantity of training samples is still not enough for LSMI to cover all the clutter rank and it can only make the maximum clutter residual 2 dB below the actual target. However, the performance for SR-STAP with the same training samples yields a 10 dB clutter suppression. Finally, Fig. 5 (c) shows the corresponding results with 40 training samples, in which both LSMI and SR-STAP could effectively suppress the clutter (the maximum clutter residual is about 12 dB below the actual target) and make the target visible from the surrounding clutter environment.

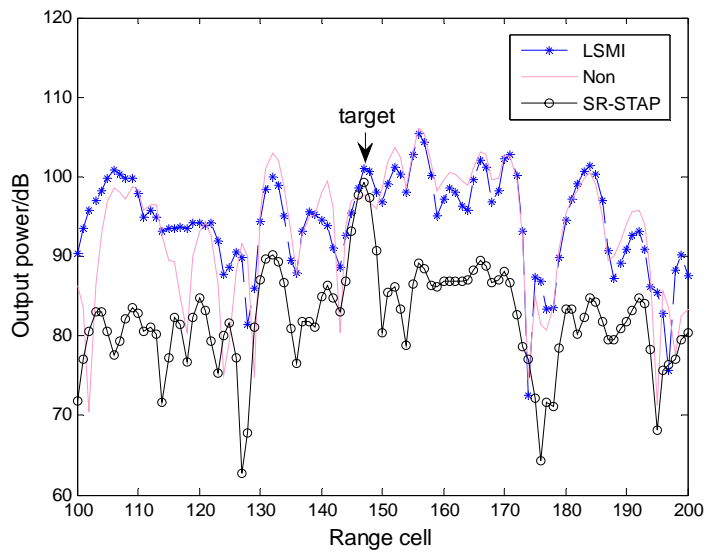

Fig. 5 (a) Range plot of Mountaintop data with 6 training samples

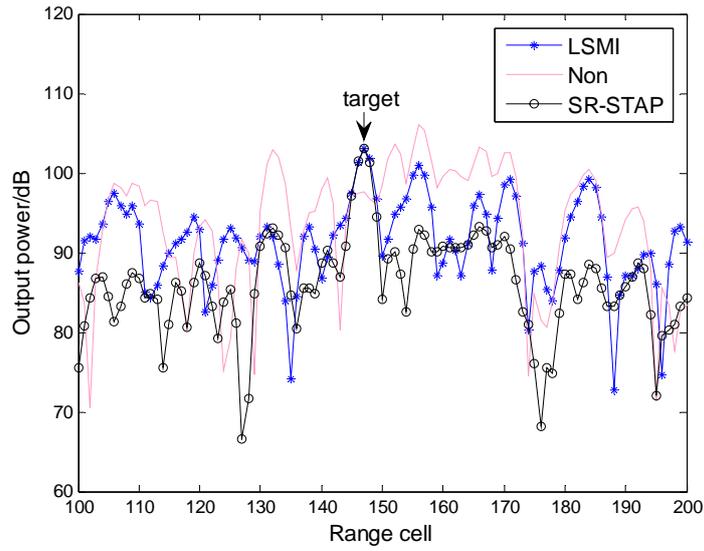

Fig. 5 (b) Range plot of Mountaintop data with 12 training samples

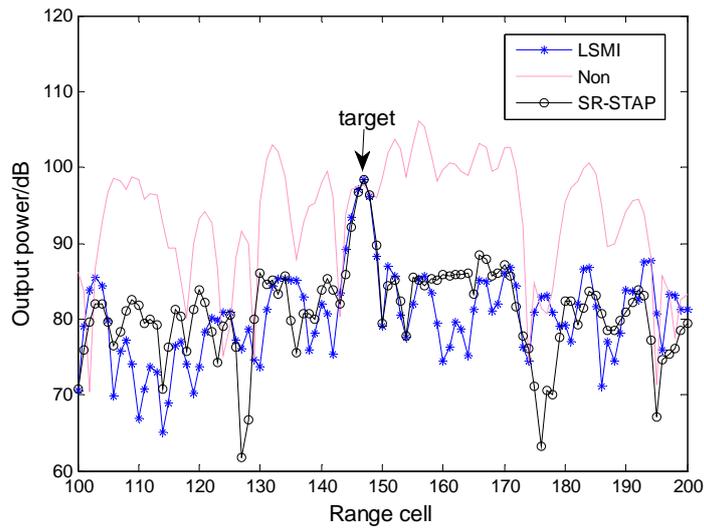

Fig. 5 (c) Range plot of Mountaintop data with 40 training samples

**4.2 Simulation experiments**

In the actual clutter scenario, where the clutter is only locally stationary, the CCM estimation with fast convergence rate has a great advantage because it can avoid the problem of the heterogeneity in the training samples [2, 5]. Therefore, the simulations in this subsection are presented to evaluate the convergence rate for CCM estimation. In addition, the influence of the mismatch in the prior knowledge is also considered because both the SR-STAP and KB-STAP approaches utilize this information in some form. Normally, the prior knowledge includes both the radar parameters and the scattering properties of the actual clutter scenario [9-10]. For simplicity, we focus only on the mismatch of the radar parameters by assuming that the scattering power is uniform, $E\{|\gamma_i|^2\} = 1, 1 \leq i \leq N_c$. The simulated scenario employs an airborne radar system incorporating a side-looking uniform linear array. The azimuthal extent of the clutter distribution is between $30° \sim 50°$, the desired target is located at azimuth $10°$ with a radial velocity $45 m/s$, and other parameters are given in Table I. All the convergence plots are based on 100 Monte Carlo simulations.

Table I Simulated parameters

| Parameter | Symbol | Value |
|---|---|---|
| Number of sensors | $N$ | 8 |
| Number of pulses | $M$ | 8 |
| Platform velocity | $v$ | 300 m/s |
| Pulse repetition interval | $PRI$ | 0.25 ms |
| Radar wavelength | $\lambda$ | 0.3 m |
| Inter-sensor spacing | $d$ | 0.15 m |
| Input clutter-to-noise ratio | $CNR$ | 35 dB |

Conventionally, the efficiency of the STAP filter is evaluated by the improvement factor (IF), which is defined as the ratio of the SCR between the output and input:

$$IF = \frac{SCR_{out}}{SCR_{in}} = \frac{|\mathbf{w}^H\mathbf{s}|^2 / \mathbf{w}^H\mathbf{R}\mathbf{w}}{\mathbf{s}^H\mathbf{s}/tr(\mathbf{R})}, \quad (32)$$

where the adaptive filter is given as $\mathbf{w} = \mu\hat{\mathbf{R}}^{-1}\mathbf{s}$, $\hat{\mathbf{R}}$ is the CCM estimation using a given technique (such as LSMI, CL or SR-STAP), $\mathbf{s}$ denotes the steering vector of the moving target, $\mathbf{R}$ is the actual CCM, and $tr(\mathbf{R})$ is the input clutter power. The IF performance is compared to the optimum value to evaluate the convergence rate as

$$IF_{Loss} = \frac{IF}{IF_{opt}} = \frac{|\mathbf{w}^H\mathbf{s}|^2}{\mathbf{w}^H\mathbf{R}\mathbf{w} \cdot \mathbf{s}^H\mathbf{R}^{-1}\mathbf{s}} \quad (33)$$

where $IF_{opt}$ is the performance given by the optimal filter $\mathbf{w} = \mu\mathbf{R}^{-1}\mathbf{s}$. The number of IID snapshots required to yield $IF_{Loss}$ within -3 dB is called the convergence rate of the processor

[2].

Next, the convergence performance of different STAP approaches is evaluated in the various clutter scenarios. The colored loading factor $\beta_d$ in CL is set to one, and $\beta_L$ is set to match the actual noise floor for all these STAP approaches. In SR-STAP, the noise allowance $\varepsilon$ is set to $10^{-4}$ and the sparsity $s$ is unknown, but can be estimated from the prior knowledge using (14)-(16). Therefore, any mismatch in the prior knowledge will cause a performance loss because both the SR-STAP and KB-STAP approaches utilize prior knowledge implicitly or explicitly. Next, we will analyze the influence of the mismatch in these parameters on both the SR-STAP and KB-STAP algorithms.

**A. Velocity mismatch**

In the first scenario, the velocity mismatch is considered and other radar parameters are kept in accordance with the actual scenario. Figure 6 gives the $IF_{Loss}$ performance versus the number of IID snapshots. The convergence rate of LSMI is slow at about 12 (twice the clutter rank) because it does not use any prior knowledge. For the CL algorithm, when the assumed velocity $v_{ass}$ coincides with the actual scenario $v_{ass}=300m/s$, the assumed CCM in (11) is identical to the actual one, so the convergence rate is exactly one. However, when there is some velocity mismatch, such as $v_{ass}=285m/s$, which is common airborne radar systems, the assumed clutter ridge will deviate from the actual one such that the assumed CCM could hardly contain all the actual clutter scatters along the actual clutter ridge. As a consequence, the convergence rate in CL will decrease to about 6. This indicates that the convergence rate of KB-STAP is sensitive to the velocity mismatch. On the other hand, SR-STAP could effectively accelerate the convergence rate in the ideal case where $v_{ass}=300m/s$ because it has the capacity of obtaining the spectral characteristics even with a few snapshots. Furthermore, the slope deviation of the clutter ridge causes little sparsity variation in the discretized angle-Doppler plane. For example, velocity deviations as great as $50m/s$ only cause a sparsity variation as $\Delta\hat{s}=1$, according to (14)-(16). Hence, SR-STAP still has enough sparsity to cover the actual clutter distribution and the spectrum estimation fits with the actual clutter scenario, per $\|\mathbf{x}-\mathbf{\Psi\alpha}\|_2 \leq \varepsilon$, in the case of velocity mismatch. More details are given to analyze the reason for this robustness.

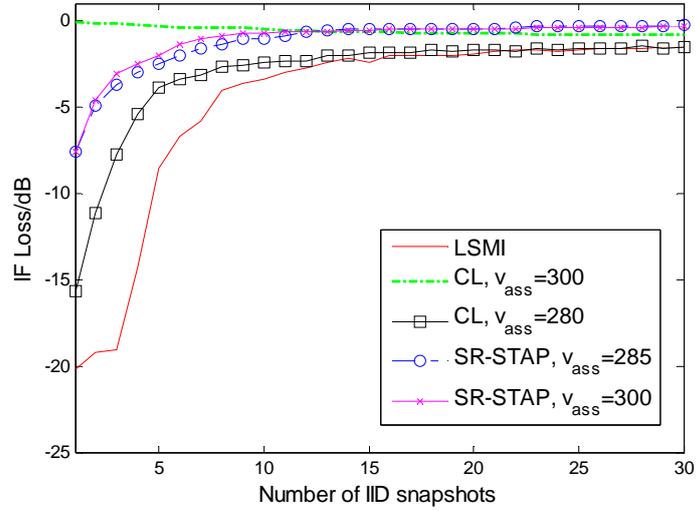

Fig. 6 $IF_{Loss}$ versus number of IID snapshots with velocity mismatch

Figure 7 gives the $IF_{Loss}$ result versus the assumed velocity with 3 IID snapshots. It is shown that CL could achieve the optimal performance when the assumed velocity $v_{ass}$ matches with the actual value, $v_{ass} = 300 m/s$. Any velocity mismatch will make the assumed clutter ridge deviate from the actual one because the velocity determines the slope of the clutter ridge. Therefore, any estimation error will cause a sharp decrease in the corresponding performance, whether it is an overestimation or underestimation. On the other hand, SR-STAP can achieve good performance at nearly all assumed velocities because little sparsity change is introduced by the velocity variation. The estimated spectrum can still fit with the actual clutter distribution as long as the sparsity is enough to cover the actual clutter distribution. In other words, although SR-STAP is influenced by the velocity variation implicitly, its spectrum estimation is mainly determined by the data samples and the algorithm is robust to the velocity mismatch.

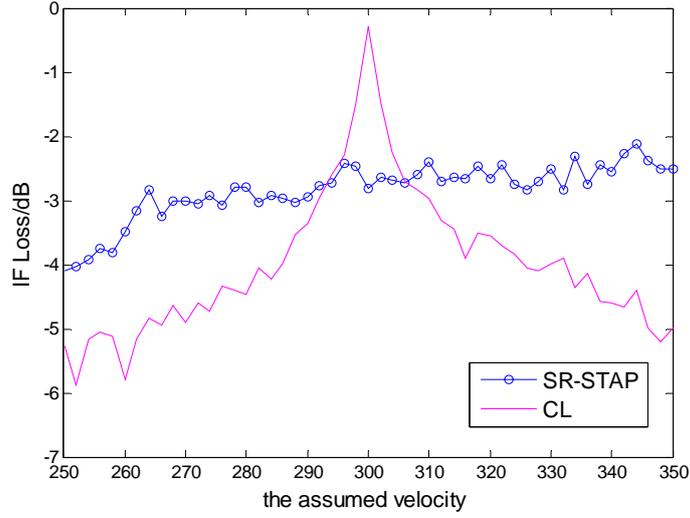

Fig. 7 $IF_{Loss}$ versus the assumed velocity

**B. Azimuth angle mismatch**

In the second scenario, the clutter azimuth center is set as $40°$ to match the actual scenario and the width mismatch is considered. Other simulated parameters are kept the same as those specified in Table I. The $IF_{Loss}$ performance versus the number of IID snapshots is given in Fig. 8 with width mismatch. LSMI does not need any prior knowledge and the convergence rate is the same as the rate shown in Fig. 6. The assumed CCM in CL is identical to the actual one when the assumed azimuth width $width_{ass}$ coincides with the actual value, $width_{ass} = 20°$, such that the convergence rate is one sample. However, when the width is overestimated, for example $width_{ass} = 40°$, CL will make extra clutter in the vicinity of the desired moving target, which is located at azimuth $10°$ with the radial velocity $45 m/s$. Therefore, the adaptive filter will force the nulls near the target and cause an $IF_{Loss}$ degradation. Furthermore, the performance improves more slowly as the number of snapshots increases because the moving target is still submerged in the extra clutter. However, SR-STAP can achieve very good performance in this case, as explained in the following section. The sparsity estimation in (14)-(16) increases in the case of width overestimation. Therefore, the joint position set $\hat{\Gamma}$ contains nearly all the positions along the actual clutter ridge, as well as some false positions, which are caused by the pseudo-peaks. The projection amplitude of the measurement $\mathbf{X}$ onto the false positions is quite small because LS minimizes the distance between the measurement $\mathbf{X}$ and the subspace $\mathbf{\Psi}_{\hat{\Gamma}}$. In this way, SR-STAP does not make extra clutter near the moving target and the clutter spectrum still fits with the actual distribution.

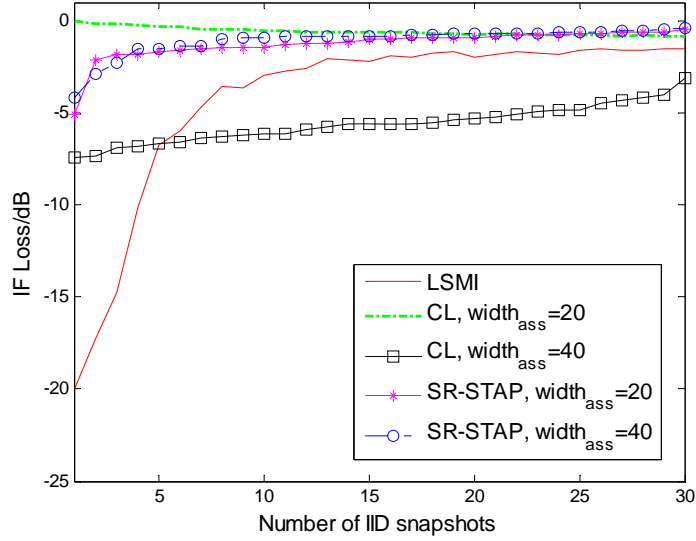

Fig. 8 $IF_{Loss}$ versus number of IID snapshots with azimuth width mismatch

Figure 9 gives the $IF_{Loss}$ versus the assumed azimuth width with 3 IID snapshots. It is shown that CL can achieve the optimal performance when the assumed width matches the actual value, $width_{ass} = 20°$. However, this performance will degrade to some extent in the case of underestimation because the assumed distribution is not wide enough and there is still some clutter residual after adaptive processing. At the same time, CL will make extra false clutter near the moving target in the case of overestimation and the corresponding $IF_{Loss}$ will degrade. SR-STAP obtains similar results in the case of underestimation because the sparsity is not enough to cover the actual clutter distribution. However, it can achieve very good performance in the case of overestimation because the amplitudes on the false positions are effectively suppressed by LS. Therefore, the CCM estimation in SR-STAP maintains the merit of fast convergence rate in the case of azimuth overestimation.

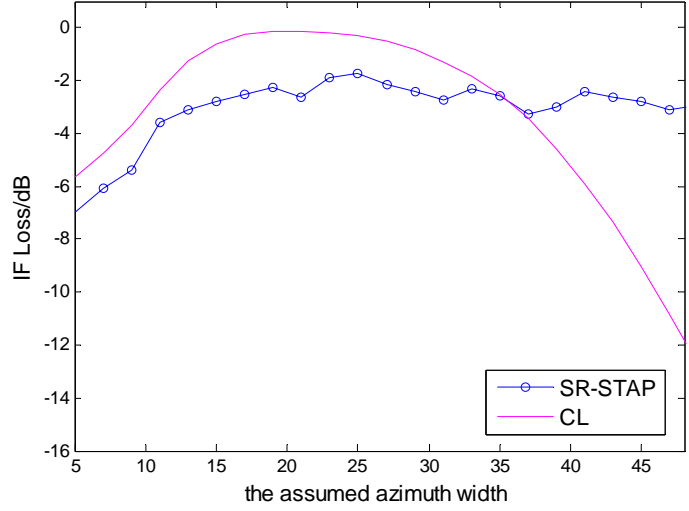

Fig. 9 $IF_{Loss}$ versus the assumed azimuth width

## C. Crab angle mismatch

In an actual airborne radar system, due to the effect of aircraft deviation, the angle-Doppler dependence of the clutter is changed into [10]

$$f_d = \frac{2v}{\lambda}\sin(\theta_s + \phi_a), \qquad (34)$$

where $\phi_a$ stands for the crab angle, i.e., the angle between the directions of the moving platform and the linear array. Therefore, the sparsity estimate in SR-STAP should be reconsidered because the Doppler spreading is related with both the azimuth angle $\theta_s$ and crab angle $\phi_a$. Here we simply recalculate the Doppler cells to consider this effect as

$$\Delta M = \left\lceil \frac{2v \cdot \left(\sin(\theta_{high} + \phi_a) - \sin(\theta_{low} + \phi_a)\right)}{\lambda \cdot PRF} \cdot \alpha_d M \right\rceil, \qquad (35)$$

so the sparsity estimation can be carried out similarly to (16). The actual crab angle is set as $\phi_a = 2°$ and the effect of mismatch is considered. Other radar parameters are kept in accordance with the actual scenario in Table I. Fig. 10 gives the $IF_{Loss}$ performance versus the number of IID snapshots for crab angle mismatch. LSMI does not need any prior knowledge and the performance remains the same. For CL, the assumed CCM is identical to the actual one and the convergence rate is one sample when the assumed crab angle $\phi_{a,ass}$ coincides with the actual value, $\phi_{a,ass} = 2°$. The assumed clutter ridge will deviate from the actual scenario when the assumed crab angle mismatches with the actual value, for example, $\phi_{a,ass} = 0°$. In this case, the convergence rate declines to about 4. However, similar to the velocity mismatch, the crab angle mismatch only impacts the slope of the assumed clutter ridge and has little influence on the corresponding sparsity. Therefore, the spectrum estimation in SR-STAP can still fit with the actual scenario even in the case of crab angle mismatch.

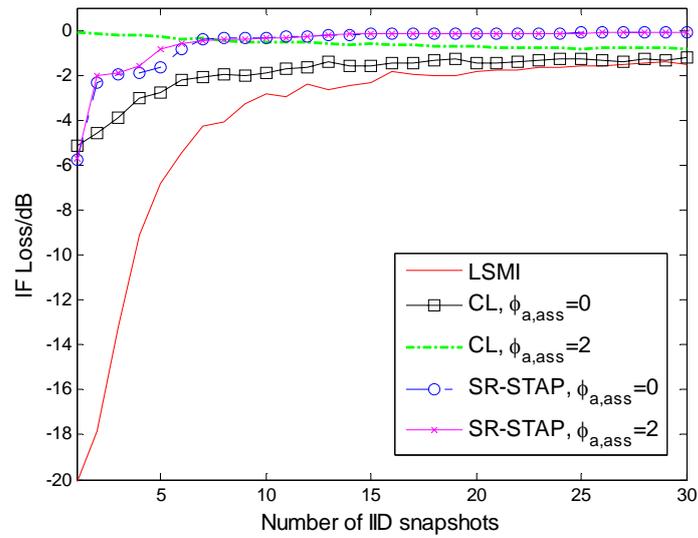

Fig. 10 $IF_{Loss}$ versus number of IID snapshots with crab angle mismatch

Fig. 11 gives the $IF_{Loss}$ versus the assumed crab angle with 3 IID snapshots. It is shown that CL can achieve optimal performance in the ideal case. Nevertheless, any estimation error will cause a sharp decrease in the corresponding performance, whether it is an overestimation or underestimation. On the other hand, although the crab mismatch makes the assumed clutter ridge deviate from the actual one, the occupied cells by the assumed clutter ridge remain nearly the same. Therefore, SR-STAP still has enough sparsity to cover the actual clutter ridge and achieves quite good performance even in the case of crab angle mismatch.

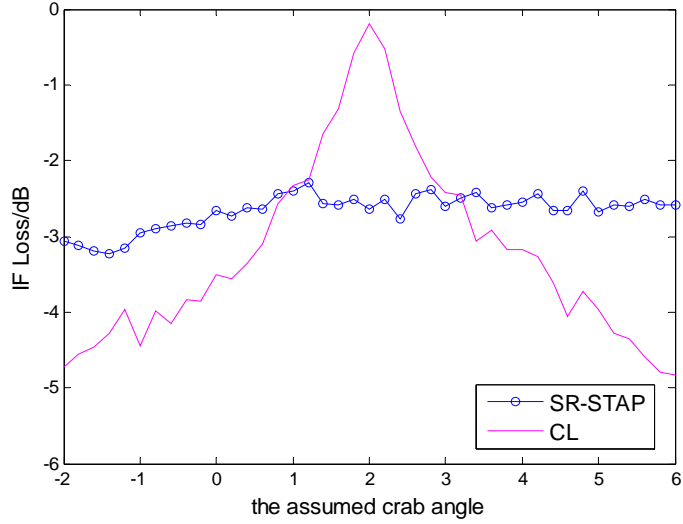

Fig. 11 $IF_{Loss}$ versus the assumed crab angle

Finally, some concluding remarks are presented for the different STAP approaches discussed above. Conventional statistical-based STAP like LSMI is directly data-based; thus it is robust, but converges slowly. At the same time, KB-STAP methods like CL can effectively accelerate the convergence rate by incorporating the prior knowledge explicitly, but these methods lack robustness. The reason for this is that CL contains both the assumed $\mathbf{R}_c$ and data-based $\hat{\mathbf{R}}_{SMI}$ parts in the CCM estimation. When the prior knowledge mismatches the actual clutter scenario, there will be some clutter residual which must be estimated by the SMI part. As a result, the convergence rate of CL will be slowed down. However, SR-STAP can obtain an accurate clutter space-time spectrum to estimate the CCM with a fast convergence rate. Moreover, SR-STAP depends less on the prior-knowledge than KB-STAP methods, so it is more robust to any mismatch between the prior knowledge and the actual clutter environment. In this way, SR-STAP can still preserve the benefits of fast convergence, even in the case of prior knowledge mismatch.

## 5. Conclusions

In this paper, we have analyzed the sparsity of the clutter distribution in the angle-Doppler domain and proposed a new SR-STAP algorithm to estimate the clutter space-time spectrum via sparse recovery. The CCM estimate as well as the STAP filter can be obtained with a fast convergence rate based on this estimated spectrum. The Mountaintop results have shown that SR-STAP can obtain an accurate clutter space-time spectrum and the corresponding filter is quite effective, such that it can suppress the maximum clutter residual to 8-12 dB below the actual target using only a few training samples. The quantitative analysis in the simulated experiments illustrated that SR-STAP can obtain the CCM estimate with a convergence rate of about 3-4 training samples, which is much faster than the conventional statistical-based STAP techniques like LSMI. Furthermore, the simulation has also shown that SR-STAP depends less on the prior knowledge, making it more robust to the mismatch of the prior knowledge than knowledge-based STAP techniques. Therefore, SR-STAP has the advantages of fast convergence rate and robustness, and therefore has great potential in actual clutter scenarios.

Some considerations for further research are given as follows. First, sparse recovery, as described in this paper, is currently based on convex optimization. This approach is stable but has a very high computational load. Therefore, it would be valuable to study the performance of the greedy methods [17-18], which are suitable for real-time processing. Second, it is also essential to consider practical nonideal factors, such as clutter internal motion or channel mismatch, in the process of sparse recovery. Consideration of these factors can improve the spectrum estimation in actual clutter scenarios.